\documentclass[aps,prb,twocolumn,10pt,amsmath,amssymb,showpacs]{revtex4-1}
\usepackage{graphicx}
\usepackage{dcolumn}
\usepackage{bm}
\usepackage{longtable}

\begin{document}
\title{Magnetization-based assessment of
correlation energy in canted Single-Chain Magnets}
\author{A.~Barasi\'{n}ski,$^{1}$ G.~Kamieniarz,$^{2}$ and A.~Drzewi\'{n}ski$^{1}$}
\affiliation{$^{1}$Institute of Physics, University of Zielona G\'ora, ul. Prof. Z. Szafrana 4a,
65-516 Zielona G\' ora, Poland}
\affiliation{$^{2}$A. Mickiewicz University, Faculty of Physics, ul. Umultowska 85, 61-614 Pozna\'n, Poland}

\date{\today}

\begin{abstract}
We demonstrate numerically that for the strongly anisotropic homometallic $S=2$ canted single-chain magnet
described by the quantum antiferromagnetic Heisenberg model the correlation energy and exchange coupling
constant can be directly estimated from the in-field-magnetization profile found along the properly selected crystallographic direction. In the parameter space defined by the spherical angles ($\phi$, $\theta$) determining
the axes orientation, four regions are identified with different sequences of the characteristic field-dependent
magnetization profiles representing the antiferromagnetic, metamagnetic and weak ferromagnetic type behavior.
These sequences provide a criterion for the applicability of the anisotropic quantum Heisenberg model to a given
experimental system.  Our analysis shows that the correlation energy decreases linearly with field and vanishes for
a given value $H_{cr}$ which defines a special coordinates in the metamagnetic profile relevant for the zero-field
correlation energy and magnetic coupling. For the single-chain magnet formed by the strongly anisotropic manganese(III)
acetate meso-tetraphenylporphyrin complexes coupled to the phenylphosphinate ligands, the experimental metamagnetic-type
magnetization curve in the $c$ direction yields an accurate estimate of the values of correlation energy
$\Delta_{\xi}/k_B = 7.93 $~K and exchange coupling $J/k_B=1.20$~K.
\end{abstract}


\maketitle

\section{Introduction}
\label{sec:int}

Molecular magnets make a wide emerging class of new materials whose properties extend the range of those
typically associated with magnets, i.e. they offer low density, transparency, electrical insulation, and
a possibility of low-temperature synthesis \cite{Dante}. Single-Chain Magnets (SCM) belonging to
the class of molecular magnets, have been of particular
interest since Gatteschi et al. \cite{Gatteschi} discovered slow relaxation of magnetization
in a chain compound, comprising Co(II) centers and organic radicals, without any evidence of phase
transition to a three-dimensional magnetic ordering.  Soon after, Clerac et al. discovered similar
magnetic properties in an $S=3$ Heisenberg ferromagnetic chain comprising Mn(III)-Ni(II)-Mn(III) trimers
with an easy axis parallel to the chain \cite{Clerac}. Next, other SCM systems often based
on ferrimagnetic chains containing alternating spins of unequal magnitude were discovered \cite{ferri}.
The slow magnetic relaxation arises from large uniaxial magnetic anisotropy, negligible magnetic interactions
between the chains and considerable intrachain interactions \cite{Dante,Coulon}. The last condition is desirable
for raising the blocking temperature. Moreover, for some one-dimensional quantum spin systems
with competing nearest-neighbor and next-nearest-neighbor interactions \cite{Ivanov} the Single-Chain Magnetic
behavior can be affected with frustration \cite{Zheng}.

The simplest description of ferromagnetic SCMs is based on the Glauber theory if the limit of an Ising chain can
be explored \cite{Glauber}. The Ising ferromagnetic chains display slow relaxation of magnetization and the relaxation
time $\tau$ depends on the spin-pair correlations. For these chains the correlation length $\xi$ diverges exponentially
at low temperatures as
\begin{equation}
\label{eq:ksi}
\xi = C_0 \exp(\Delta_{\xi}/k_BT),
\end{equation}
where $\Delta_{\xi}$ is the correlation energy and $k_B$ is the Boltzmann
constant \cite{Miyasaka}. The exponent $\Delta_{\xi}$ is proportional to the activation energy which enters
the Arrhenius law \cite{Miyasaka} and can be directly estimated from the experimental static susceptibility of
the chain by the relation $\xi \sim \chi T$, plotting ln($\chi T$) as a function of 1/$T$. The correlation energy
is also referred to as the energy to create a domain wall. Important fact is that $\Delta_{\xi}$ is proportional to
the magnetic coupling constant \cite{Miyasaka} and similar conclusions remain valid for the antiferromagnetic Ising
model \cite{Coulon,Gloria} but then $\xi \sim 1/(\chi T)$. In addition, the correlation energy $\Delta_{\xi}$ is
also the main part of the additive term determining the relaxation time in the Ising-like and anisotropic classical
Heisenberg chains \cite{Miyasaka,Sessoli}.

The relations between the correlation length and the zero-field susceptibility as well as between the correlation
energy and the magnetic coupling can be used for the classical ferromagnetic Heisenberg model in
the strong anisotropy limit \cite{Miyasaka,Barbara,Loveluck}. If the Hamiltonian is defined by
$${\cal H} = J \sum_{i=1}^{L-1} \left( S_{i}^{x}S_{i+1}^{x} + S_{i}^{y}S_{i+1}^{y} + S_{i}^{z}S_{i+1}^{z} \right) + 
D \sum_{i=1}^{L} S_{i}^zS_{i}^z,$$
the limit is reached when the exchange coupling $J$ and the single-ion anisotropy $D$ satisfy the relation
$|D/J| > 2/3$. Then the correlation energy calculated from the product $\chi T$ yields the value of
the coupling $J$ because \cite{Miyasaka} $$\Delta_{\xi} = 2|J|S^2.$$

For ferromagnetic systems, this scenario is well confirmed \cite{Miyasaka,Coulon2004,Saitoh} and has been exploited
for a number of compounds \cite{Boeckmann,Zheng47}, but for antiferromagnetic chains with non-collinear anisotropy
axes \cite{Sessoli,Jung} the situation is much less clear. The problem was revealed \cite{Sessoli} for the compound
of formula [Mn(TPP)O$_2$PHPh]$\cdot $H$_2$O considered a textbook example of SCM (TPP = {\it meso}-tetraphenylporphyrin
and PHPh = phenylphosphinate) and referred to as Mn-CAF (canted antiferromagnet). It was possible to rationalise many
experimental results performed on Mn-CAF, assuming the generalised expression for the correlation energy in the form
 \begin{equation}
\label{eq:DeltaKsi}
\Delta_{\xi} = 2|J|S^2 {\mit \cos} \delta,
\end{equation}
where $\delta$ is the canting angle for two classical spins oriented along the anisotropy axes. However,
the proportionality between the correlation length $\xi$ and the product $\chi T$ was ruled out, implying that
the correlation energy cannot be determined by the magnetic measurement of the static susceptibility.

In this paper we demonstrate that the correlation energy of the quantum model of a canted antiferromagnetic chain
can be found from the in-field single-crystal magnetization profile properly chosen and its $\delta$ dependence is
given by the classical expression (\ref{eq:DeltaKsi}). These results suggest that the correlation energy $\Delta_{\xi}$
can be estimated from the in-field magnetization measurements. As soon as the spin value, the geometrical structure
and $\Delta_{\xi}$ are known, the value $J$ of the magnetic coupling follows from Eq. (\ref{eq:DeltaKsi}). Considering
the empirical shape of the magnetization isotherm in the $c$ direction for Mn-CAF, we directly find the $J$ value
without any fitting procedure. 

The paper is organized as follows: Sec.~\ref{sec:1} is devoted to a model describing the anisotropic quantum
spin systems and the numerical method for accurate calculations of magnetic
properties in a wide temperature range. Sec.~\ref{sec:2} is dedicated to presentation of results
and their discussion. Sec.~\ref{sec:3} concludes our paper, summarizing the main outcomes.

\section{Model and DMRG method for canted Single-Chain Magnets}
\label{sec:1}

We consider the quantum anisotropic Heisenberg model which is needed to get quantitative estimates of the thermodynamic
properties \cite{BDK,SBKD} of the chains with non-collinear anisotropy axes
\begin{multline} 
\label{eq:01}
{\cal H} = J \sum_{i=1}^{L-1} \left( S_{i}^{a}S_{i+1}^{a} + S_{i}^{b}S_{i+1}^{b} + S_{i}^{c}S_{i+1}^{c} \right) +\\
\sum_{i=1}^{L} \sum_{\alpha\beta} \left( S_{i}^{\alpha}\hat{D}_{i}^{\alpha\beta}S_{i}^{\beta} + \mu_{B}H^{\alpha}\hat{g}_{i}^{\alpha\beta}S_{i}^{\beta} \right),
\end{multline}
\noindent
where spin $S = 2$ and $L$ stands for the length of the chain. The exchange coupling $J$ between nearest-neighbor
spins is isotropic, uniform and positive. The tensors representing the single-ion anisotropy $\hat{D}_{i}$
and $\hat{g}_{i}$-factors are non-diagonal and depend on angles ($\phi$, $\theta$) or ($- \phi$, $\theta$) for odd
and even sites, respectively. The indices $\alpha,\, \beta\, \in \, \{a, b, c\}$ define the global coordination system.
The $c$ direction is chosen along the chain axis and plays also the role of the quantization axis. 

\vspace{2mm}
\begin{figure}[tbh]
\centering
\includegraphics[scale=0.35]{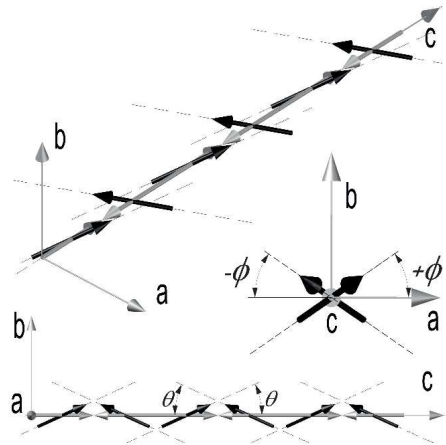}
\caption{Schematic views of the chain structure in the global coordination system $\{a, b, c\}$, where
the spherical angles ($\phi$, $\theta$) are defined. The black arrows represent the canted Mn-CAF spin arrangement,
whereas the gray ones correspond to the collinear antiferomagnetic spin arrangement ($\phi=0$, $\theta=0$).
}
\label{fig:00}
\end{figure}

The explicit form of the tensors in the model (\ref{eq:01}) depends on the angles ($\phi$, $\theta$) defining
the anisotropy axes (Fig.~\ref{fig:00}) and is given in the previous publications \cite{Sessoli,BDK}. We note
that for all the pairs of angles ($\phi$, $\theta$) the canting angle is determined from the relation
$$\cos\delta = \sin^2(\theta) \cos(2 \phi)+ \cos^2(\theta).$$

Analysis of our model (\ref{eq:01}) is based on the numerical Quantum Transfer Matrix
method \cite{Kam1,Caramico,Kam2,Matysiak,SBKD}, where the partition function of the quantum chain is mapped onto
the partition function of the classical $2d$ system with multi-spin interactions and a finite
width $2M$ \cite{Suzuki2,Delica}. For different values of $M$, called the Trotter number, the classical
partition functions form a series of approximants $Z_M$, where the leading errors are of the order of $1/M^2$.
The higher the value of $M$, the better the quantum nature of the problem is taken into account. As the Hilbert
space dimension increases exponentially with increasing Trotter number, computations are feasible only for
relatively small $M$. To overcome this problem, we employ the Density-Matrix Renormalization Group
technique (DMRG) \cite{dmrg,Wang,Drzew,Drzew2,Shibata} for determining an effective Hamiltonian representation
in the Hilbert subspace of a given size $m$ which ensures covering the entire experimental temperature range.

The DMRG method is very powerful but it is computationally demanding as far as the resources and programming are
concerned. The transfer matrices are asymmetric and the corresponding density matrices, constructed from the right
and left eigenvectors of the largest eigenvalue $\lambda_{max}$ of a transfer matrix, are non-Hermitian. We were able
to overcome these problems by applying complex algebra \cite{Artur}, which guaranties the biorthogonality of
complex eigenvectors but increases the computational complexity.

In some DMRG applications \cite{Wang} the Hilbert space is split into subspaces of fixed values of
$M_{stag}=\sum (-1)^j S_j$, where $S_j$ denotes a spin variable in a given column. Then the highest eigenvalue
$\lambda_{max}$ can be found in the block with the staggered magnetization $M_{stag}=0$. This simplification
follows from some symmetries which may occur in the four-spin DMRG vertices. For the non-collinear model this is
not the case, therefore $\lambda_{max}$ has to be found in the entire Hilbert space.

For each temperature, the free energy of the macroscopic system (\ref{eq:01}) per site ($L \to \infty$) is
related to the maximum eigenvalue $\lambda_{max}$ of the transfer matrix ${\cal T}_M$ \cite{Baxter},
whereas the magnetization and magnetic susceptibility can be obtained by taking the first and second derivatives
of the free energy with respect to external magnetic field. These quantities are calculated in
the subspace $0 \leq \phi \leq \pi/2$ and $0 \leq \theta \leq \pi/2$ which follows from the Hamiltonian symmetry.

The correlation length of the $\alpha$ components of spin is given by the following ratio \cite{Suzuki}:
\begin{align} \label{eq:02}
1/\xi_{\alpha} &= -\lim_{R \to \infty} \frac{1}{R} \log \langle S_i^{\alpha}S_{i+R}^{\alpha} \rangle
\\
&= -\lim_{M \to \infty} \lim_{R \to \infty} \lim_{L \to \infty} \frac{1}{R}
\log \frac {{\rm Tr} {\cal T}_M^i S^{\alpha} {\cal T}_M^R S^{\alpha} {\cal T}_M^{L-(i+R)}} 
{{\rm Tr} {\cal T}_M^{L}}
\notag \\
&= \lim_{M \to \infty} \log \frac{\lambda_{max}}{\lambda_{\nu}}, \notag
\end{align}
where $S^{\alpha}$ is the tensor product of the ${\alpha}$-component of the spin operator and $2M-1$
identity matrices of $2S+1$ size. The $| \psi_{max} \rangle$ vector corresponds to the $\lambda_{max}$
eigenvalue, whereas  the eigenvalue $\lambda_{\nu}$ is the highest among all whose eigenvector
$| \psi_{\nu} \rangle$ which satisfy the condition $|\langle \psi_{max}|S^{\alpha}|\psi_{\nu} \rangle | \neq 0$.

The convergence of the approximants with respect to $M$ depends on temperature and the size of the optimal basis
set $m$. Our results are provided for $m= 125$ and the Trotter number $M$ up to $20$ (the lower temperature
the higher $M$ is needed). According to our estimation the accuracy of results in the whole parameter space
considered is not lower than 0.5\% \cite{SBKD}. 

\section{Results and discussions}
\label{sec:2}
Recently several compounds of antiferromagnetic one-dimensional systems containing Mn(III) ions have been
reported \cite{Sessoli,Jung}. Here the most interesting example \cite{Sessoli} is Mn-CAF, where the anisotropy
axes  are defined by Jahn-Teller elongation of Mn(III) octahedra and make the angle $\theta=21.01$ deg with the chain
direction parallel to the $c$ axis (see Fig.~\ref{fig:00}). Projections of the anisotropy axes onto the plane
perpendicular to the chain, alternate along the $a$ crystallographic axis making angles $\phi=56.55$ deg.

The single-crystal magnetometry measurements \cite{Sessoli} revealed three patterns of the in-field-magnetization
profiles. For the $a$ direction the pattern shows linear field dependence and is referred to as antiferromagnetic (AF).
The field dependencies measured along the $b$ and $c$ are referred to as the weak ferromagnetic (WF) and
the metamagnetic type (MM), respectively.

Taking into account the known values of $\phi$ and $\theta$ angles and performing the Monte Carlo simulations, 
the values of exchange coupling, uniaxial anisotropy and $g$-factor were obtained (in short the Mn-CAF parameters)
for the classical counterpart of the model (\ref{eq:01}),  
\begin{equation}
\label{eq:00}
J/k_B=1.36(8)\,K,\,D/k_B=-4.7(2)\,K,\,g=1.97(1),
\end{equation}
fitting the experimental susceptibility and magnetization curves \cite{Sessoli}. So, the high value of
the ratio $|D/J| \simeq 3.5$ corresponding to a strong anisotropy occurred. 

Using the classical Mn-CAF parameters (\ref{eq:00}) and the corresponding angles
$$\phi = 56.55\; \mbox {deg,}\; \theta=21.01\; \mbox {deg,}\; \delta=34.6\; \mbox {deg},$$
we have determined the in-field magnetization profiles for three crystallographic directions. Their features agree
qualitatively with those published previously in Fig.~5 for Mn-CAF \cite{Sessoli}, including the sequence of patterns
AF, WF and MM for the $a$, $b$, $c$ directions. The $MM$-type profile for the field along the $c$ axis was calculated
earlier within DMRG \cite{BDK,GK} and its behavior detected some quantitative deviations with respect to
the classical counterpart.

\vspace{2mm}
\begin{figure}[tbh]
\centering
\includegraphics[scale=0.3]{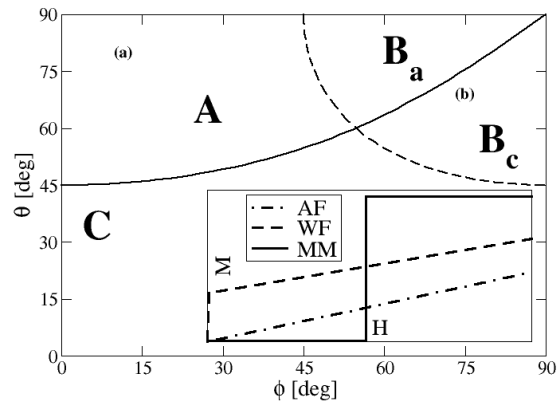}
\caption{
The angle-dependent diagram showing sectors with different sequences of the in-field magnetization patterns.
In sectors A, B, C the $MM$-type patterns occur in the $a$, $b$, $c$ directions, respectively. All the patterns
are outlined in the inset in the zero-temperature limit.}
\label{fig:03}
\end{figure}

Relaxing the angles $\phi$ and $\theta$, we have noticed that the sequence of the magnetization patterns varies with
the easy axes orientation and we have established the domains (Fig.~\ref{fig:03}) in the ($\phi$,\;$\theta$)
parameter space. Within a given domain the fixed shape of the low-temperature magnetization isotherms as a function
of field for all the crystallographic directions occurs. The expected zero-temperature patterns are plotted in
the inset. The sectors appearing in the diagram are described in Table~\ref{tab:01}. In sectors A, B, C the $MM$-type
patterns are unveiled for the field along the $a$, $b$, $c$ axes, respectively. The sequence of the AF and WF profiles
accompanying the $MM$-type shape in a given sector is specified in Table~\ref{tab:01}. In the B sectors the AF pattern
is missing and two WF patterns are present in return. The crystallographic direction, where the magnetization values
are higher in the low fields, has been distinguished by an additional index $a$ (or $b$).

\begin{center}
\begin{longtable}{|c|c|c|c|}
\caption{Pattern sequences in the diagram sectors.} \label{tab:01} \\
\hline
\multicolumn{1}{|c|}{Sector} &  \multicolumn{3}{|c|}{Crystallographic direction}\\
\endfirsthead

\cline{2-4}
symbol &~~~~ $a$ ~~~~&~~~~ $b$ ~~~~& $c$ \\
\hline

A & $MM$ & $WF$ & $AF$  \\
B$_a$ & $WF$ & $MM$ & $WF$  \\
B$_c$ & $WF$ & $MM$ & $WF$  \\
C & $AF$ & $WF$ & $MM$  \\
\hline
\end{longtable}
\end{center}

The B sectors in the diagram are separated from the other sectors by the dashed line obtained from the solution of
the equation $ \cot(2 \theta) = \sin(2 \phi) − \cos(2 \phi)$. This line corresponds to the canted angle between
the adjacent anisotropy axes which amounts to $\delta = \pi/2$. The solid line in the diagram follows from the symmetry
of the Hamiltonian and can be determined from the solution of the equation $\cot\theta = \cos\phi$.

The diagram in Fig.~\ref{fig:03} provides the criterion whether the model (\ref{eq:01}) can describe a particular
magnetic system. It predicts the occurrence of the $MM$-type profile which is accompanied by a given combination
of the $AF$-type and the $WF$-type patterns.

We immediately see that the criterion is fulfilled for Mn-CAF \cite{Sessoli}. Its geometric coordinates ($\phi$, $\theta$)
belong to the sector C and the experimental single-crystal profiles agree with the patterns predicted for this sector.

\vspace{2mm}
\begin{figure}[tbh]
\centering
\includegraphics[scale=0.33]{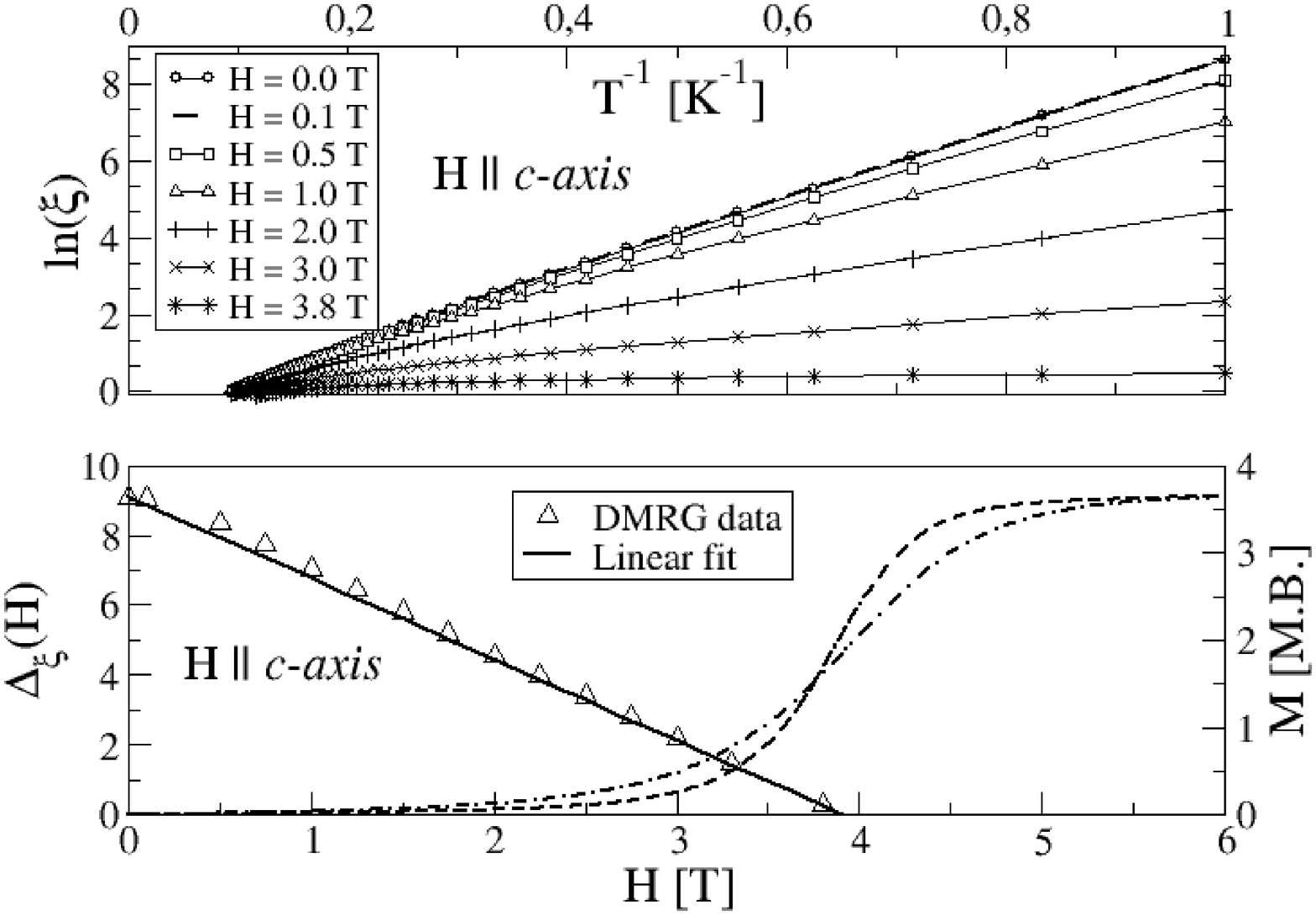}
\caption{(Above) Temperature behavior of the correlation length for various values of the
magnetic field $H$ parallel to the $MM$ axis (the Mn-CAF parameters).
(Bellow) Our estimations of the $\Delta_{\xi}(H)$ vs external magnetic field parallel to the $c$ axis
plotted by triangles. The solid line represents the linear regression: $f(H) = 9.10 - 2.33*H$.
The magnetization isotherms at $T = 1$ and $1.6$ K are given by the dashed and chain lines, respectively.}
\label{fig:04}
\end{figure}

As the zero-field susceptibility and the correlation energy are not related for the canted antiferromagnetic chains,
we have decided to analyse the field dependence of the correlation energy $\Delta_{\xi}(H)$. First we checked if
the correlation length diverges in the limit of low temperatures. For low temperatures the upper panel of
Fig.~\ref{fig:04} demonstrates a linear dependence of $\ln(\xi_c)$ (where $\xi_c$ is determined from Eq.~(\ref{eq:02}))
on the inverse of temperature, choosing the Mn-CAF parameters (\ref{eq:00}) and the corresponding angles $\phi = 56.55$
and $\theta = 21.01$. This dependence implies that the low-temperature correlation length $\xi$  exhibits
the exponential divergence in the $MM$-direction
\begin{eqnarray}
\label{eq:03}
\xi = C_0 \exp(\Delta_{\xi}(H)/k_BT),
\end{eqnarray}
but the slope $\Delta_{\xi}(H)$ is field dependent.

The field dependence of the correlation energy $\Delta_{\xi}(H)$ has been studied numerically and plotted by
triangles in the bottom part of Fig~(\ref{fig:04}). We have revealed the striking linear behavior which can be
very well reproduced by the function 
\begin{eqnarray}
\label{eq:f}
\Delta_{\xi}(H)/k_B = 9.10 - 2.33*H,
\end{eqnarray}
where the right hand side is expressed in K and the field is given in T.
So, the function $\Delta_{\xi}(H)$ vanishes, which implies a finite value of correlation length $\xi$
(see Eq.~(\ref{eq:03})), for the critical field $H_{cr} = 3.91$~T, what is close to the value estimated directly
from the DMRG correlation length, $H_{cr} = 3.86$~T.

The abrupt change of the correlation length from the infinite to the finite value induced by the field should be
accompanied by some changes in the physical properties. We demonstrate in the bottom panel of Fig.~\ref{fig:04} that
the $MM$-type profile displays a rapid jump in the vicinity of $H_{cr}$~T from $0$ to a nearly saturation value.
We interpret this behavior as a field-driven transition from the antiferromagnetic to the ferromagnetic order along
the $c$ direction. Such critical field can be determined from the intersection of the low-temperature magnetization
isotherms, from their inflection point or from the maximum of the in-field magnetic susceptibility.

\begin{figure}[tbh]
\centering
\includegraphics[scale=0.3]{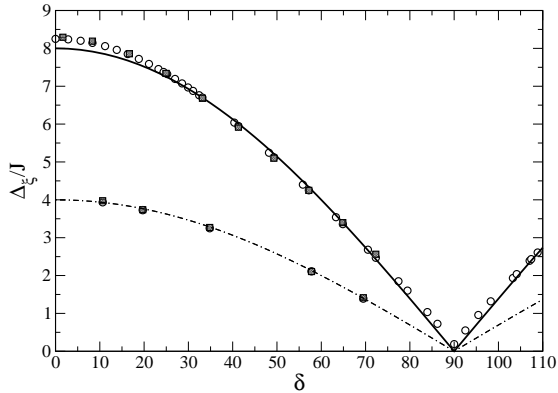}
\caption{Comparison between the functions $2S^2|\cos(\delta)|$ and $S^2|\cos(\delta)|$ $vs \; \delta$ plotted by the full and the dashed line, respectively, and the ratios $\Delta_{\xi}/J$ and $\mu_B M_{cr} H_{cr}/J$ illustrated by the circles and squares. The open symbols correspond to $J/k_B = 1.26$ K and the full symbols to $J/k_B = 1.36$ K.
} 
\label{fig:05}
\end{figure}

Now we analyze the correlation energy $\Delta_{\xi}$ for the Mn-CAF parameters in the absence of a magnetic field.
The formula (\ref{eq:f}) implies $\Delta_{\xi}(H=0)/k_B=9.10$ K, while the DMRG calculations based on
Eqs. (\ref{eq:ksi}) and (\ref{eq:02}) yield $\Delta_{\xi}/k_B=9.05$ K. These estimates are in remarkable
agreement with the classical prediction \cite{Sessoli} $\Delta_{\xi}=8.95$ K following from Eq. (\ref{eq:DeltaKsi}).
Moreover, the coincidence of the correlation energy of the quantum model (\ref{eq:01}) and the classical dependence
(\ref{eq:DeltaKsi}) is confirmed and demonstrated for a number of couplings $J$ and the canting angles $\delta$
in Fig.~\ref{fig:05}.

\vspace{6mm}

In Fig.~\ref{fig:05} the upper continuous and lower dashed lines represent the functions 2S$^2 \cos \delta$ and
S$^2 \cos \delta$, respectively. The open and full circles illustrate the ratios $\Delta_{\xi}/J$ calculated
numerically within DMRG for $J/k_B$=1.26~K and $J/k_B$=1.36~K, respectively. As to the angles $\delta$, their values
follow from the angles $\phi$ and $\theta$ chosen from different sectors of the diagram in Fig.~\ref{fig:03}.
The fields were applied in the proper direction for a sector selected.

We conclude that the agreement between the classical predictions 2S$^2 \cos \delta$ for the ratio $\Delta_{\xi}/J$ and
those calculated for our quantum model (\ref{eq:01}) and plotted in Fig.~\ref{fig:05} is very good. There are some
deviation for small and high values of $\delta$, but they can be attributed to uncertainties of the extrapolations
performed for ln($\xi$) yielding the values $\Delta_{\xi}$. These uncertainties are of the order of a few percent
and are much higher than the accuracy of the DMRG results. They could be diminished lowering temperature and increasing
the cost of our simulations. Concluding, our numerical results in Fig.~\ref{fig:05} give a strong evidence that
the correlation energy of the canted antiferromagnetic quantum chains can be expressed by
the relation (\ref{eq:DeltaKsi}) previously found for the classical systems.

In Fig.~\ref{fig:05} we also plot by the full squares the ratios $\mu_B M_{cr} H_{cr}/J$ as a function of $\delta$,
where $M_{cr}$ is the value of the $MM$-type magnetization profile corresponding to the field $H_{cr}$. We emphasise
that the ratio is completely determined by the coordinates of the inflection point of the $MM$-type magnetization
profile. The ratios $\mu_B M_{cr} H_{cr}/J$ calculated coincide with the values of the function S$^2 \cos \delta$ drawn
by the dashed line and are lower by a factor of 2 than the ratios $\Delta_{\xi}/J$, implying that
\begin{eqnarray}
\label{eq:MH}
\Delta_{\xi} = 2 \mu_B M_{cr} H_{cr}.
\end{eqnarray}
We can interpret the result (\ref{eq:MH}) that the correlation energy is equal to the average Zeeman energy per pair
of spins calculated for the critical field $H_{cr}$.

The relation (\ref{eq:MH}) is an important result of our paper because it provides a direct estimate of the correlation
energy $\Delta_{\xi}$ in terms of the special coordinates of the $MM$-type magnetization profile. Moreover,
the exchange constant $J/k_B$ can be also obtained from the $MM$-type single-crystal low-temperature magnetization
isotherm. Combining Eqs. (\ref{eq:DeltaKsi}) and (\ref{eq:MH}), we find the formula for the magnetic coupling  
\begin{eqnarray}
\label{eq:J}
J = \mu_B M_{cr} H_{cr}/(S^2 \cos \delta).
\end{eqnarray}

\begin{figure}[tbh]
\centering
\includegraphics[scale=0.35]{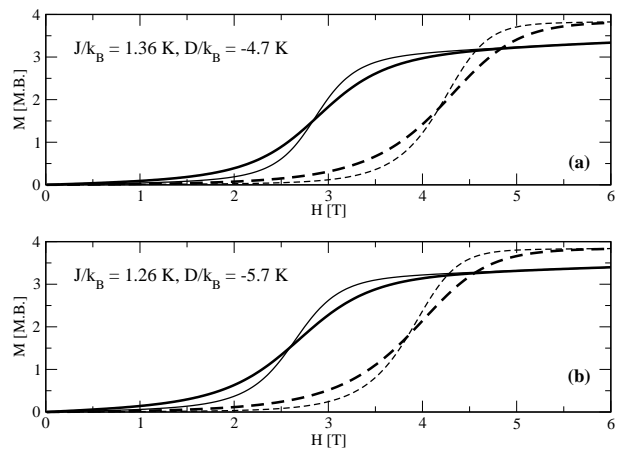}
\caption{Metamagnetic profiles as a function of magnetic field for two sets of parameters presented in the legends.
The thick lines correspond to $T=1.6$ K and the thin lines to $T=1.0$ K. The solid and the dashed lines belong to
the $A$ sector ($\phi = 10$ deg, $\theta = 80$ deg) and to the $B_C$ sector ($\phi = 75$ deg, $\theta = 65$ deg),
respectively.}
\label{fig:02b}
\end{figure}

To validate our conclusions, in Fig.~\ref{fig:02b} the $MM$-type behavior is drawn for the ratios $|D/J|  \simeq 3.5$
and $|D/J|  \simeq 4.5$ in both panels. For the angles specified in the legends and coming from the sectors A and B,
the a and b panels display the profiles found for the magnetic field parallel to the $a$ axis and $b$ axis,
respectively. The slope of all the $MM$-type curves in Fig.~\ref{fig:02b} increases with decreasing temperature
and the shape does not change.

The coordinates of the intersections of the corresponding magnetization isotherms in Fig.~\ref{fig:02b}
($T=1.0$ K and $1.6$ K) inserted into Eq.~(\ref{eq:J}) yield the consistent estimates of $J/k_B$ with the uncertainty
equal to $\pm 0.04$ K for the input values $J/k_B = 1.26$ K and $1.36$ K. When we used lower temperatures ($T=0.5$ K
and $1.0$ K), the uncertainty declined to $\pm 0.005$ K for the same $J/k_B$, which is the expected trend.

\vspace{6mm}

Now, taking the values $\delta=34.6$ deg, $H_{cr}=3.45$ T and $M_{cr}=1.71$ (in $\mu_B$) directly from the experiment
for Mn-CAF compound \cite{Sessoli}, we can evaluate from Eqs.~(\ref{eq:MH}) and (\ref{eq:J}) the value of correlation
energy $\Delta_{\xi}/k_B = 7.93$ K and the coupling constant $J/k_B=1.20$ K, respectively. Both the values are slightly
lower than the $\Delta_{\xi}/k_B=8.95$ K and $J/k_B=1.36(8)$ K reported earlier \cite{Sessoli}. However, this fact can
be understood as the experimental value $J/k_B=1.36(8)$ K was settled on the basis of the classical model and
the fitting procedure which usually overestimate the coupling constant \cite{SBKD}. If the coupling $J$ is
overestimated then the relation (\ref{eq:DeltaKsi}) may proliferate the inaccuracy and imply somewhat
higher $\Delta_{\xi}$.

\section{Conclusions}
\label{sec:3}

Our investigation has confirmed that the relative positions of the anisotropy axes affect the shape of magnetization
curves along different crystallographic directions. The resulting angle-dependent diagram
predicts four areas, each characterized by a specific sequence of the in-field magnetization patterns. If for
a compound studied the shapes of magnetization curves in the low-temperature range are consistent with those
established in the diagram for a given geometry, then the compound can be analyzed in terms of the quantum
Heisenberg model with a strong anisotropy.

For the canted antiferromagnet SCM the values of correlation energy $\Delta_{\xi}(H)$ decrease linearly with
the applied field and vanish at $H_{cr}$ which coincides with the intersection point of the $MM$-type magnetization
profiles. The coordinates $H_{cr}$ and $M_{cr}$ of this special point determine the zero-field correlation energy
$\Delta_{\xi}$ which in turn reveals the classical dependence on the canting angle $\delta$ and the coupling $J$.
These findings relate the $J$ value to the coordinates $H_{cr}$ and $M_{cr}$ and yield a simple recipe to
the evaluation of $J$.

Following these ideas, we have estimated for the Mn-CAF compound the relevant quantities at
$\Delta_{\xi}/k_B = 7.93$ K and $J/k_B = 1.20$ K which agree very well with the values reported earlier.

Our calculations provide some insight into the common claim that in the strong anisotropy limit the Heisenberg model
can be substituted by the Ising model. We confirm that the projections of spins onto the axis displaying the $MM$-type
pattern can be treated as classical Ising-type variables. Then the rapid jump in the magnetization profile can be
related to formation of an Ising domain wall. Anyway, it is striking that the quantity $\Delta_{\xi}$ defined in
the absence of field can be so easily settled from the experiment performed in the presence of field.

\section{Acknowledgements}

We are grateful to Prof. R. Sessoli for many valuable discussions explaining the structure and properties of
the compound considered here. A.~B. and G.~K. thank to the Polish Ministry of Science and Higher Education for
supporting this work under the Iuventus Plus program (IP2010 001170) and Grant Nr 230137, respectively. 
Numerical calculations were performed in PSNC Pozna\'n (Poland) and WCSS Wroc\l aw (Poland, grant 82). This
work was also granted access to the HPC resources within DECI by the PRACE-2IP under grant no RI-283493.


\begin{thebibliography}{99}

\bibitem{Dante} Dante~Gatteschi, Roberta~Sessoli, and Jacques~Villain, {\it Molecular nanomagnets},
Oxford University Press, 2006.
\bibitem{Gatteschi} A.~Caneschi, D.~Gatteschi, N.~Lalioti, C.~Sangregorio, R.~Sessoli, G.~Venturi,
A.~Vindigni, A.~Rettori, M.~G.~Pini, M.~A.~Novak, Angew. Chem. Int. Ed. {\bf 40}, 1760 (2001).
\bibitem{Clerac}  R.~Clerac, H.~Miyasaka, M.~Yamashita, C.~Coulon,  J.~Am.~Chem.~Soc. {\bf 124}, 12837 (2002).
\bibitem{ferri} H.~Miyasaka, T.~Madanbashi, K.~Sugimoto, Y.~Nakazawa, W.~Wernsdorfer, K.~Sugiura,
M.~Yamashita, C.~Coulon, R.~Clerac, Chem.~Eur.~J. {\bf 12}, 7028 (2006).
\bibitem{Coulon} C.~Coulon, H.~Miyasaka, and R.~Cle\'{e}rac, Struct.~Bond. {\bf 122}, 163 (2006);
H.~Miyasaka and M.~Yamashita, Dalton Trans., 399 (2007), and references therein.
\bibitem{Ivanov} N.~B.~Ivanov, J.~Richter, and U.~Schollwoeck, Phys.~Rev.~B {\bf 58}, 14456 (2007).
\bibitem{Zheng} Y.-Z.~Zheng, W.~Xue, W.-X.~Zhang, M.-L.~Tong, X.-M.~Chen, F.~Grandjean, G.~J.~Long, S.-W.~Ng,
P.~Panissod, and M.~Drillon, Inorg.~Chem. {\bf 48}, 2028 (2009).
\bibitem{Glauber} R.~J.~Glauber, J.~Math.~Phys. {\bf 4}, 294 (1963).
\bibitem{Miyasaka} H.~Miyasaka, M.~Julve, M.~Yamashita, and R.~Clerac, Inorg.~Chem. {\bf 48}, 3420 (2009).
\bibitem{Gloria} M.~G.~Pini and A.~Rettori, Phys.~Rev.~B {\bf 76}, 064407 (2007).
\bibitem{Sessoli} K.~Bernot, J.~Luzon, R.~Sessoli, A.~Vindigni, J.~Thion, S.~Richeter, D.~Leclercq,
J.~Larionova, and A.~van~der~Lee, J.~Am.~Chem.~Soc. {\bf 130}, 1619 (2008).
\bibitem{Barbara} B.~Barbara, J.~Phys. {\bf 34}, 1039 (1973).
\bibitem{Loveluck} J.~M.~Loveluck, S.~W.~Lovesey, S.~J.~Aubry, Phys.~C:~Solid~State~Phys. {\bf 8}, 3841 (1975).
\bibitem{Coulon2004} C.~Coulon, R.~Clerac, L.~Lecren, W.~Wernsdorfer, and H.~Miyasaka
Phys.~Rev.~B {\bf 69}, 132408 (2004).
\bibitem{Saitoh} A.~Saitoh, H.~Miyasaka, M.~Yamashita, and R.~Clerac, J.~Mater.~Chem. {\bf 17}, 2002 (2007).
\bibitem{Boeckmann} J.~Boeckmann, M.~Wriedt, C.N\"ather, Chem. Eur. J. {\bf 18}, 5284 (2012).
\bibitem{Zheng47} Y-Z.~Zheng, W.~Xue, M-L.~Tong, X-M.~Chen, F.~Grandjean, G.J.~Long, Inorg. Chem. {\bf 47}, 4077 (2008)
\bibitem{Jung} Jung~Hee~Yoon, Dae~Won~Ryu, Hyoung~Chan~Kim, Sung~Won~Yoon,
Byoung~Jin~Suh, and Chang~Seop~Hong, Chem.~Eur.~J. {\bf 15}, 3661 (2009).
\bibitem{BDK} A.~Barasi\'{n}ski, G.~Kamieniarz, and A.~Drzewi\'{n}ski, Comp.~Phys.~Comm.
{\bf 182}, 2013 (2011).
\bibitem{SBKD} P.~Sobczak, A.~Barasi\'{n}ski, G.~Kamieniarz, A.~Drzewi\'{n}ski, Phys.~Rev.
{\bf B 84}, 224431 (2011).
\bibitem{Kam1} G.~Kamieniarz, R.~Matysiak, A.~C.~D'Auria, F.~Esposito, and U.~Esposito, Phys. Rev.
{\bf B 56}, 645 (1997).
\bibitem{Caramico} A.~Caramico~D'Auria, F.~Esposito, U.~Esposito, D.~Gatteschi, G.~Kamieniarz, and S.~Wa\l cerz,
J. Chem. Phys. {\bf 109}, 1613 (1998).
\bibitem{Kam2} G.~Kamieniarz, M.~Bieli\'{n}ski, and J.-P.~Renard, Phys.~Rev. {\bf B 60}, 14521 (1999).
\bibitem{Matysiak}  R.~Matysiak, G.~Kamieniarz, P.~Gegenwart, and A.~Ochiai, Phys.~Rev. {\bf B 79}, 224413 (2009).
\bibitem{Suzuki2} M.~Suzuki, Prog.~Theor.~Phys. {\bf 56}, 1454 (1976).
\bibitem{Delica} T.~Delica and H.~Leschke, Physica {\bf A 176}, 736 (1990).
\bibitem{Wang} X.~Q.~Wang and T.~Xiang, Phys.~Rev. {\bf B 56}, 5061 (1997).
\bibitem{dmrg} U.~Schollwoeck, Rev.~Mod.~Phys. {\bf 77}, 259 (2005); K.~Hallberg, Adv.~Phys. {\bf 55},
477 (2006); U.~Schollwoeck, Ann.~Phys.~(NY) {\bf 326}, 96 (2011).
\bibitem{Drzew} A.~Macio\l ek, A.~Drzewi\'{n}ski, and R.~Evans, Phys.~Rev. {\bf B 64}, 056137 (2001).
\bibitem{Drzew2} A.~Drzewi\'{n}ski, A.~Macio\l ek, A.~Barasi\'{n}ski, and S.~Dietrich,
Phys.~Rev. {\bf E 79}, 041145 (2009).
\bibitem{Shibata} N.~Shibata, J.~Phys.~A:~Math.~Gen. {\bf 36}, R381 (2003).
\bibitem{Artur} A.~Barasi\'{n}ski, P.~Sobczak, A.~Drzewi\'{n}ski, G.~Kamieniarz, A.~Bie\'{n}ko,
J.~Mrozi\'{n}ski, and D.~Gatteschi, Polyhedron {\bf 29}, 1485 (2010).
\bibitem{Baxter} R.~J.~Baxter, {\it Exactly Solved Models in Statistical Mechanics}, Academic Press 1982.
\bibitem{Suzuki} N.~Hatano and M.~Suzuki, J.~Phys.~Soc.~J. {\bf 62}, 1346 (1993).
\bibitem{GK} G.~Kamieniarz, P.~Kozlowski, M.~Antkowiak, P.~Sobczak, T.~\'{S}lusarski, D.~M.~Tomecka,
A.~Barasi\'{n}ski, B.~Brzostowski, A.~Drzewi\'{n}ski, A.~Bie\'{n}ko, J.~Mrozi\'{n}ski,
Acta Phys. Pol. {\bf A 121}, 992 (2012).
\end{thebibliography}
\end{document}